\documentclass[fleqn]{an}
\usepackage{graphicx}
\usepackage{times}
\usepackage{bm}
\Pagespan{43}{50}
\Yearpublication{2011}
\Yearsubmission{2010}
\Month{9}
\Volume{332}
\Issue{1}
\DOI{10.1002/asna.201012345}
\begin{document}
\sloppy

\newcommand{\EQ}{\begin{equation}}
\newcommand{\EE}{\end{equation}}
\newcommand{\EQA}{\begin{eqnarray}}
\newcommand{\EEA}{\end{eqnarray}}
\newcommand{\brac}[1]{\langle #1 \rangle}
\newcommand{\pd}{\partial}
\newcommand{\pdz}{\partial_z}
\newcommand{\DIV}{\vec{\nabla} \cdot }
\newcommand{\CURL}{\vec{\nabla} \times }
\newcommand{\cross}[2]{\boldsymbol{#1} \times \boldsymbol{#2}}
\newcommand{\crossm}[2]{\brac{\boldsymbol{#1}} \times \brac{\boldsymbol{#2}}}
\newcommand{\ve}[1]{\boldsymbol{#1}}
\newcommand{\mean}[1]{\overline{#1}}
\newcommand{\meanv}[1]{\overline{\bm #1}}
\newcommand{\cst}{c_{\rm s}^2}
\newcommand{\nut}{\nu_{\rm t}}
\newcommand{\etat}{\eta_{\rm t}}
\newcommand{\etatz}{\eta_{\rm t0}}
\newcommand{\memf}{\overline{\bm{\mathcal{E}}}}
\newcommand{\memfi}{\overline{\mathcal{E}}_i}
\newcommand{\etaT}{\eta_{\rm T}}
\newcommand{\urms}{u_{\rm rms}}
\newcommand{\Urms}{U_{\rm rms}}
\newcommand{\brms}{B_{\rm rms}}
\newcommand{\Beq}{B_{\rm eq}}
\newcommand{\eu}{\hat{\bm e}}
\newcommand{\xu}{\hat{\bm x}}
\newcommand{\yu}{\hat{\bm y}}
\newcommand{\zu}{\hat{\bm z}}
\newcommand{\Ou}{\hat{\bm \Omega}}
\newcommand{\kef}{k_{\rm f}}
\newcommand{\tauc}{\tau_{\rm c}}
\newcommand{\tauto}{\tau_{\rm to}}
\newcommand{\HP}{H_{\rm P}}
\newcommand{\St}{{\rm St}}
\newcommand{\Sh}{{\rm Sh}}
\newcommand{\Pm}{{\rm Pm}}
\newcommand{\Rm}{{\rm Rm}}
\newcommand{\Pra}{{\rm Pr}}
\newcommand{\Ra}{{\rm Ra}}
\newcommand{\Ma}{{\rm Ma}}
\newcommand{\Tay}{{\rm Ta}}
\newcommand{\Ro}{{\rm Ro}}
\newcommand{\Rey}{{\rm Re}}
\newcommand{\Co}{{\rm Co}}
\newcommand{\Cost}{\Omega_\star}
\newcommand{\ReLS}{{\rm Re}_{\rm LS}}
\newcommand{\qxx}{Q_{xx}}
\newcommand{\qyy}{Q_{yy}}
\newcommand{\qzz}{Q_{zz}}
\newcommand{\qxy}{Q_{xy}}
\newcommand{\qxz}{Q_{xz}}
\newcommand{\qyz}{Q_{yz}}
\newcommand{\qij}{Q_{ij}}
\newcommand{\Omx}{\Omega_x}
\newcommand{\Omz}{\Omega_z}
\newcommand{\emf}{\bm{\mathcal{E}}}
\newcommand{\emfi}{\mathcal{E}_i}
\newcommand{\nab}{\mbox{\boldmath $\nabla$} {}}
\newcommand{\meanFFFF}{\overline{\mbox{\boldmath ${\cal F}$}}{}}{}
\def\onethird{{\textstyle{1\over3}}}
\def\onehalf{{\textstyle{1\over2}}}
\def\threefourths{{\textstyle{3\over4}}}

\title{On global solar dynamo simulations}

\author{P.J. K\"apyl\"a\inst{1,2}\fnmsep\thanks{Corresponding author:
    {petri.kapyla@helsinki.fi}}}

\titlerunning{On global solar dynamo simulations}
\authorrunning{P.J. K\"apyl\"a}

\institute{ 
Department of Physics, PO BOX 64 (Gustaf H\"allstr\"omin katu 2a), 
FI-00014 University of Helsinki, Finland
\and
NORDITA, AlbaNova University Center, Roslagstullsbacken 23, SE-10691 
Stockholm, Sweden}

\received{2010 Aug 31} 
\accepted{2010 Nov 10}
\publonline{2011 Jan 12}


\abstract{%
  Global dynamo simulations solving the equations of
  magnetohydrodynamics (MHD) have been a tool of astrophysicists who
  try to understand the magnetism of the Sun for several decades
  now. During recent years many fundamental issues in dynamo theory
  have been studied in detail by means of local numerical simulations
  that simplify the problem and allow the study of physical effects in
  isolation. Global simulations, however, continue to suffer from the
  age-old problem of too low spatial resolution, leading to much lower
  Reynolds numbers and scale separation than in the Sun. Reproducing
  the internal rotation of the Sun, which plays a crucual role in the
  dynamo process, has also turned out to be a very difficult
  problem. In the present paper the current status of global dynamo
  simulations of the Sun is reviewed. Emphasis is put on efforts to
  understand how the large-scale magnetic fields, i.e.\ whose length
  scale is greater than the scale of turbulence, are generated in the
  Sun. Some lessons from mean-field theory and local simulations are
  reviewed and their possible implications to the global models are
  discussed. Possible remedies to some current issues of solar
  simulations are put forward.}
\maketitle

\section{Introduction}
\label{sec:intro}
The large-scale magnetic field of the Sun varies quasiperiodically in
time and space: the amount of sunspots varies with an average period
of 11 years whereas the period of the magnetic field itself is 22
years. Sunspots appear on a latitude strip $\pm40$ degrees away from
the equator, with spots appearing at high latitudes in the beginning
of the cycle and progressively closer to the equator as the cycle
advances. Explaining this activity has been one of the principal goals
of solar physicists since the first detection of magnetic fields in
the Sun by Hale (1908) and Hale et al.\ (1919).

Nowadays it is generally accepted that the magnetic field of the Sun
is maintained by a dynamo residing within, or just below, the
convection zone which occupies roughly the outer third of solar
radius. The problem is that the flows within the solar convection zone
are highly turbulent. Especially in the early days of dynamo theory,
computational capabilities were very limited rendering direct
solutions of the MHD equations impossible.

The first successful models of solar magnetism were based on a
statistical description of turbulent eddies under the influence of
rotation and their interaction with large-scale shear (Parker
1955). Work along similar lines evolved into a rigorous mathematical
theory, now often referred to as turbulent mean-field dynamo theory,
where the separation of small and large scales plays a crucial role
(e.g.\ Moffatt 1978; Parker 1979; Krause \& R\"adler 1980; R\"udiger
\& Hollerbach 2004). In turbulent mean-field dynamo theory,
large-scale magnetic fields are maintained by the combined action of
helical turbulence ($\alpha$-effect) and large-scale shear flow
against turbulent diffusion. Mean-field models cabaple of reproducing
the main features of solar observations have existed for decades
(e.g.\ Parker 1955; Steenbeck \& Krause 1969; K\"ohler 1973; see
Ossendrijver 2003 for a recent review).

Mean-field models rely on parametrizations of turbulence, such as the
$\alpha$-effect and turbulent diffusion, that we refer to as turbulent
transport coefficients. In the absence of observational data or
methods to extract them from direct numerical simulations, the
turbulent transport coefficients had to be computed from ill-defined
approximations. Such procedure often involves a number of free
parameters that can be tuned in the mean-field models, which is
obviously not a satisfactory state of affairs. Only very recently has
an efficient method for computing turbulent transport coefficients
from simulations surfaced in the form of the so-called test-field
method (Schrinner et al.\ 2005, 2007).

As the computing power increased, attempts to model solar magnetism by
solving the equations of magnetohydrodynamics directly, started to
surface (e.g.\ Gilman 1983; Glatzmaier 1985). More sophisticated
simulations have continued to appear ever since (e.g.\ Brun et al.\
2004; Browning et al.\ 2006; Ghizaru et al.\ 2010). However, none of
the current models can reproduce the main features of solar magnetic
activity (see, e.g.\ Miesch \& Toomre 2009). Another aspect that the
simulations still struggle with is the internal rotation of the Sun:
most simulations produce angular velocity profiles that are dominated
by the Taylor--Proudman balance and cylindrical isocontours whereas
the in the Sun the contours are more conical (e.g.\ Thompson et al.\
2003). Furthermore, the shear layers close to the top and at the base
of the convection zone, both of which have been suggested as the
locations of the solar dynamo (e.g.\ Parker 1993; Brandenburg 2005),
cannot yet be reproduced numerically in a self-consistent manner.

The problems that direct simulations are facing today are most likely
caused by the fact that the parameter regime that is accessible by
simulations is still too far removed from solar
conditions. Unfortunately, realistic values of the Rayleigh and
Reynolds numbers are not likely to be reached any time soon which
means that the models need to be improved in a more clever way if
progress is to be made. This could include more sophisticated
subgrid-scale models and boundary conditions, and numerical techniques
to increase resolution in places where it is most needed. In this
paper some of these issues are discussed and possible remedies are
suggested.

The paper is organised as follows: Sect.~\ref{sec:numcha} summarizes
the main numerical issues encountered in global dynamo simulations. In
Sects.~\ref{sec:MFT} and \ref{sec:local} possible guidance from
turbulent mean-field theory and local simulations are discussed,
respectively. Sections~\ref{sec:global} and \ref{sec:missing}
summarize the current state of global solar dynamo simulations and
their possible caveats. Final thoughts are given in
Sect.~\ref{sec:discuss}.

\section{Numerical challenges}
\label{sec:numcha}
Here it is assumed that stellar interiors can be dealt within the
scope of the MHD approximation and that the gas obeys the equation of
state of ideal gas. These assumptions are quite likely violated in the
very uppermost and lowermost parts of the solar convection zone where
radiation becomes important, but here we assume their effects to be
minor for large-scale dynamos. Even then a realistic model of the
solar and stellar dynamos must overcome three major numerical
challenges: (i) the small molecular diffusivities lead to immense
Rayleigh and Reynolds numbers, (ii) the time scale of thermal
relaxation is far removed from the turnover time of the turbulence,
and (iii) the convection zones of stars are extremely stratified with
more than 20 pressure scale heights. Some dimensionless parameters
relevant for the Sun are listed in Table~\ref{tab:dimpa} (see also
Ossendrijver et al.\ 2003; Brandenburg \& Subramanian 2005).

The only way to address issue (i) and to reach realistic Rayleigh and
Reynolds numbers is to radically increase the resolution of the
simulations. Given that in the Sun the fluid Reynolds number is of the
order of $10^{12}$, ${\Rey}^{3/4}\approx10^9$ grid points per
direction would be needed for all physically relevant scales to be
resolved (e.g.\ Robinson \& Chan 2001). The largest global simulations
to date can afford of the order of $10^3$ grid points per direction
(Miesch et al.\ 2008). Even if the computing power continues to
increase at the current rate, it will take decades before sufficient
resolution can be reached. Furthermore, the thermal and magnetic
Prandtl numbers ($\Pr$ and $\Pm$) are much smaller than unity, leading
to much larger lenght scales for the temperature and magnetic field
than that of the velocity. For example, in the Sun the smallest scale
of velocity is $10^7$ times smaller than that of the temperature. This
implies that numerical resolution of at least $10^7$ grid points is
needed to resolve both scales in the same model. Similar, although
somewhat less extreme, contrast is encountered with the magnetic
fields.

The second issue concerns the vastly varying time scales involved in
the solar convection zone: the turnover time of convection cells on
the surface of the Sun is of the order of minutes whereas the period
of the magnetic cycle is 22 years. However, the most severe issue is
due to the thermal relaxation (Kelvin--Helmholtz) time scale which is
of the order of a $10^7$ years for the Sun. This means that the energy
flux flowing through the convection zone is small in comparison to the
internal energy. Furthermore, this leads to a very small Mach number
in the bulk of the convection zone. In such cases the time step in the
simulations is determined by the large sound speed at the base of the
convection zone and not by the fluid velocity. This issue can be
alleviated by the use of the anelastic approximation (e.g.\ Gough
1969; Brun et al.\ 2004) which, however, breaks down near the
surface. Currently no global models are capable of dealing with both
the small Mach number flows in the deep layers and transonic flows
near the surface.

Issue (iii) arises due to the immense density stratification and leads
to similar problems as in (i): a minimum number of grid points, of the
order of five, is required to resolve a pressure scale height
$\HP$. Close to the surface of the Sun $\HP\approx100$~km so we could
get away with a grid resolution of $20$~km. Given that the depth of
the solar convection zone is $2\cdot10^5$~km, a minimum of $10^4$ grid
points is required to resolve this. Such resolution is not quite
within the grasp of simulations as of yet. However, using a
non-uniform grid in the radial direction (e.g.\ Chan \& Sofia 1986;
Robinson \& Chan 2001) can alleviate this issue.

In summary, very few parameters can have their realistic values in
global simulations (cf.\ Table~\ref{tab:dimpa}). Possibly the only
exception is the Coriolis, i.e.\ inverse Rossby, number, which spans
from roughly 10 at the base of the convection zone to $10^{-3}$ near
the surface. 
However, it is not possible to cover this range in a single model
either.
If the large-scale dynamo of the Sun is driven by a
turbulent dynamo relying on helical turbulence arising from the
interaction of rotation and stratified turbulence (see below), then it
might not be a problem that we cannot reach realistic Rayleigh and
Reynolds numbers. Currently the best hope is that as long as $\Ra$,
$\Rey$, and $\Rm$ are sufficiently high as to produce vigorous
turbulence, and the rotational influence is correctly modelled, the
main aspects of solar magnetism can be captured.

\begin{table*}
 \centering
 \caption{Summary of some dimensionless parameters in the Sun and in 
   typical simulations. The last column denotes whether the 
   simulations capture the solar regime (+) or not (--), and (--/+) 
   indicates that the range is reachable but not in a single model. 
   Here $g$ is the 
   acceleration due to gravity, $d$ is the typical scale of turbulence, 
   $\delta$ is the superadiabaticity, $\nu$ is the viscosity, $\eta$ is 
   the magnetic diffusivity, $H_{\rm P}$ is the pressure scale height,
   $u$ is a typical velocity, $\chi$ is the thermal diffusivity, whereas
   $p$ and $c_s$ are the pressure and sound speed, respectively, 
   and $\Omega$ is the rotation rate.}
\label{tab:dimpa}
\begin{tabular}{lccc}\hline
Parameter  & Sun & Simulations & Comparability \\
\hline
$\Ra=gd^4 \delta/(\nu \chi H_{\rm P})$ & $10^{20}$ & $10^7$ & --\\
$\Rey=u d/\nu$  & $10^{12}$ & $<10^4$ & -- \\
$\Rm=u d/\eta$  & $10^{9}$ & $<10^4$ & -- \\
$\Pra=\nu/\chi$ & $10^{-7}$ & $0.01$ & -- \\
$\Pm=\nu/\eta$  & $10^{-6}\ldots10^{-4}$ & $10^{-3}$ & -- \\
$N_{\rm P} = \ln(p_{\rm base}/p_{\rm top})$ & 20 & $\approx5$ & --\\
$\Ma=u/c_s$ & $10^{-4}\ldots 1$ & $10^{-4}\ldots 1$ & --/+ \\
$\delta=\nabla-\nabla_{\rm ad}$ & $10^{-8}\ldots0.1$ & $10^{-8}\ldots0.1$ & --/+ \\
$\Tay=4\Omega^2 d^4/\nu^2$ & $10^{19}\ldots10^{27}$ & $10^8$ & --\\
$\Co=2\Omega d/u$ & $10^{-3}\ldots10$ & $10^{-3}\ldots10$ & --/+\\
\hline
\end{tabular}
\end{table*}

\section{Guidance from mean-field theory}
\label{sec:MFT}
It is useful to make a small recourse into theory in order to have an
idea when a large-scale dynamo can be expected to be excited. In
mean-field dynamo theory the evolution of the large-scale magnetic
field is governed by the averaged induction equation
\begin{equation}
\frac{\pd \meanv{B}}{\pd t} = \bm\nabla\times(\meanv{U}\times\meanv{B} + 
\memf -\eta\mu_0 \meanv{J}),
\end{equation}
where the overbars denote a suitable average, $\bm{U}$, $\bm{B}$, and
$\bm{J}=\mu_0^{-1}\bm\nabla\times\bm{B}$ are the velocity, magnetic
field, and current density, respectively. Furthermore, $\eta$ is the
magnetic diffusivity and $\mu_0$ is the vacuum permeability. The extra
term in comparison to the standard induction equation is the
electromotive force
\begin{equation}
\memf=\overline{\bm{u}\times\bm{b}},
\end{equation}
where $\bm{u}=\bm{U}-\meanv{U}$ and $\bm{b}=\bm{B}-\meanv{B}$ are the
fluctuations of velocity and magnetic field, respectively. Given that
the large-scale field $\meanv{B}$ varies slowly in space and time,
$\memf$ can be written in terms of the large-scale quantities where
turbulent transport coefficients describe the effects of turbulence on
the large scales (e.g.\ Krause \& R\"adler 1980):
\begin{equation}
\memfi = \alpha_{ij}\mean{B}_j + \eta_{ijk}\frac{\pd \mean{B}_j}{\pd x_k} + \cdots,
\label{equ:memfi}
\end{equation}
where $\alpha_{ij}$ and $\eta_{ijk}$ are second and third rank
tensors, respectively, and the dots indicate the possibility to take
higher order derivatives into account.

In simple systems, such as homogeneous, isotropic turbulence, the
first term on the rhs of Eq.~(\ref{equ:memfi}) describes the
$\alpha$-effect whereas the second term is responsible for turbulent
diffusion:
\begin{equation}
\memf=\alpha \meanv{B} -\etat \mu_0 \meanv{J},
\end{equation}
where $\alpha$ and $\etat$ are scalars (Steenbeck et al.\ 1966). In
the high conductivity limit these scalars are given by
\begin{equation}
\alpha = - \onethird \tauc \mean{\bm{\omega}\cdot\bm{u}}, \quad \etat= \onethird \tauc \mean{\bm{u}^2},
\end{equation}
where $\tauc$ is the correlation time of turbulence and
$\mean{\bm{\omega}\cdot\bm{u}}$ is the kinetic helicity. In more
realistic systems $\alpha$ is no longer directly proportional to
kinetic helicity (e.g.\ R\"adler 1980), although it is still an often
used proxy. 

In the absence of shear, the $\alpha$-effect alone is able to overcome
turbulent diffusion and excite a large-scale dynamo.
This can be quantified by requiring that a dimensionless
dynamo number
\begin{equation}
D_\alpha =\frac{\alpha d}{\etat},
\end{equation}
where $d$ is the spatial extent of the system (e.g.\ the radius of the
Sun or the depth of the convection zone), exceeds a threshold value.
At the same time, the magnetic Reynolds number has to exceed a
critical value. However, for large-scale dynamos this is typically of
the order of unity and thus not an issue for the Sun (e.g.\ Krause \&
R\"adler 1980; Brandenburg 2009; K\"apyl\"a et al.\ 2010b). Another
essential ingredient is the separation of scales: the turbulent energy
carrying scale must be smaller by a factor of few in comparison to the 
system size for the large-scale dynamo to work. Numerical simulations 
in idealised setups (Brandenburg
2001) have shown such $\alpha^2$-dynamos exist but it is not likely
that this type of dynamo is the main contributor to solar
magnetism. This is because most $\alpha^2$-dynamos produce
non-oscillatory solutions, although non-uniform $\alpha$-profiles can
excite oscillatory modes as well (e.g.\ Baryshnikova \& Shukurov 1987;
R\"udiger et al.\ 2003; Stefani \& Gerbeth 2003).

When shear is present, not only is the dynamo easier to excite, but
the solutions often exhibit oscillatory solutions or dynamo waves. The
direction of propagation of such waves in $\alpha\Omega$-dynamos is
determined by the sign of the product of radial shear and the
$\alpha$-effect (e.g.\ Yoshimura 1975). According to symmetry
considerations the simplest form of the $\alpha$-effect in a rotating
stratified atmosphere of a star is given by (e.g.\ Krause \& R\"adler
1980):
\begin{equation}
\alpha_{ij} = \alpha_1 \delta_{ij} \hat{\bm{G}}\cdot\hat{\bm{\Omega}} + \alpha_2 (\hat{G}_i \hat{\Omega}_j + \hat{G}_j \hat{\Omega}_i),
\end{equation}
where $\hat{\bm{G}}$ and $\hat{\bm{\Omega}}$ denote the unit vectors
along the direction of inhomogeneity (e.g.\ turbulence intensity or
density stratification due to gravity) and rotation,
respectively. This suggests that $\alpha$ is positive (negative) in
Northern (Southern) hemisphere in the Sun. The early dynamo models
postulated (e.g.\ Parker 1955; K\"ohler 1973) a positive $\alpha$ in
the Northern hemisphere and negative radial shear within the convection
zone which produces an equatorward migrating dynamo
wave. Helioseismology, however, has revealed that the regions of
negative radial shear in the solar convection zone are situated in the
tachocline at high latitudes and in the surface shear layer in the
outermost five per cent of solar radius. The realization that it is
actually quite difficult to obtain low latitude equatorward migrating 
activity with
$\alpha\Omega$-dynamos is sometimes referred to as the `dynamo
dilemma' (Parker 1987). However, the profile and magnitude of the
$\alpha$-effect and turbulent diffusivity in the solar convection zone
are rather poorly known.

During recent years the importance of magnetic helicity conservation
has been realized in the nonlinear saturation of large-scale dynamos
(e.g.\ Brandenburg \& Subramanian 2005 and references therein). More
specifically, if magnetic field lines are confined within the object,
magnetic helicity can change only due to microscopic magnetic
diffusion leading to extremely long saturation time scales (see e.g.\
Brandenburg 2001). The Sun, however, is not a closed system and can
shed the small-scale magnetic helicity e.g.\ by coronal mass
ejections. Thus it is probably important to design simulation setups
so that magnetic helicity can escape without hindering the growth of
the large-scale magnetic fields.

\section{Lessons from comparisons of theory and local simulations}
\label{sec:local}
Early local simulations of turbulent convection failed to generate
appreciable large-scale magnetic fields (e.g.\ Nordlund et al.\ 1992;
Brandenburg et al.\ 1996) although all the necessary ingredients
(turbulence, rotation, and stratification) for an $\alpha^2$-dynamo
were present. Around the same time theoretical studies and supporting
numerical simulations suggested that generating large-scale magnetic
fields becomes all but impossible in the regime of large magnetic
Reynolds number (Cattaneo \& Vainshtein 1991; Vainshtein \& Cattaneo
1992; Cattaneo \& Hughes 1996). Furthermore, convection simulations
yielded conflicting results for the $\alpha$-effect, suggesting values
close to theoretical expectations (Brandenburg et al.\ 1990;
Ossendrijver et al.\ 2001,2002; K\"apyl\"a et al.\ 2006a) or close to
zero (Cattaneo \& Hughes 2006; Hughes \& Cattaneo 2008), further
adding to the confusion.

The early local convection simulations all lacked an important 
ingredient, namely a
large-scale shear flow. Adding sufficiently strong shear indeed
excites a large-scale dynamo (K\"apyl\"a et al.\ 2008; Hughes \&
Proctor 2009), similarly as in non-helically forced turbulence
simulations (Yousef et al.\ 2008a,b; Brandenburg et al.\ 2008).
However, the origin of the large-scale fields still remained
controversial due to the widely differing estimates of $\alpha$ (see,
e.g.\ Hughes \& Proctor 2009). Here the test-field method (Schrinner
et al.\ 2005, 2007) comes to the rescue: with it, all of the relevant
turbulent transport coefficients, including the turbulent magnetic
diffusivity, can be computed from the simulations. Performing such
analysis to the simulations of K\"apyl\"a et al.\ (2008) it turns out
that the large-scale dynamos in the presence of shear cannot be
accounted for by the $\alpha$-effect alone, but other turbulent
mean-field effects, such as the $\meanv{\Omega}\times\meanv{J}$ (R\"adler
1969,1980) and shear--current effects (Rogachevskii \& Kleeorin
2003,2004), also contribute (K\"apyl\"a et al.\ 2009b).

Similarly, the test-field results indicate that increasing the
rotation rate decreases turbulent diffusion and increases $\alpha$,
suggesting that large-scale $\alpha^2$-dynamo action becomes possible
at sufficiently rapid rotation (K\"apyl\"a et al.\ 2009b). This was
indeed realized by direct simulations in the same parameter regime
(K\"apyl\"a et al.\ 2009a; see also Jones \& Roberts 2000; Rotvig \&
Jones 2002). On the other hand, the previously obtained small values
$\alpha$ (e.g. Cattaneo \& Hughes 2006; Hughes \& Cattaneo 2008;
Hughes \& Proctor 2009) turn out to be artefacts of the so-called
imposed field method which does not take the inhomogeneities of the
large-scale field into consideration (K\"apyl\"a et al.\ 2010a, see
also Hubbard et al.\ 2009).

Another aspect that has been studied mainly using local simulations is
the nonlinear saturation of large-scale dynamos. In particular, it is
of great interest to study what happens to the saturation level of the
large-scale magnetic field when magnetic helicity fluxes are either
allowed or suppressed. It turns out that open boundaries allow
saturation on a dynamical timescale and large-scale field strengths
around equipartition with the turbulence (K\"apyl\"a et al.\
2010b). When the flux is suppressed, the large-scale field strength
decreases steeply as a function of the magnetic Reynolds number.  In
some cases this might explain why no large-scale dynamo is seen with
periodic or perfectly conducting boundaries (Tobias et al.\ 2008)
whereas the same system with magnetically open boundaries shows a
strong large-scale field (K\"apyl\"a et al.\ 2008).

The main lesson from comparisons of theory and local simulations is
that the predictions of mean-field theory need to be taken seriously:
most of the results of direct simulations can be reproduced
qualitatively and many also quantitatively by mean-field models using
turbulent transport coefficients from corresponding test-field runs
(e.g. K\"apyl\"a et al.\ 2009b; Gressel 2010). Furthermore, local
simulations have shown that the resulting large-scale dynamo is
sensitive to the magnetic boundary conditions which in many cases can
be understood in terms of magnetic helicity conservation.

\section{Global simulations}
\label{sec:global}
We consider here three classes of models that solve the equations of
magnetohydrodynamics without the mean-field approximations: forced
turbulence simulations where convection and large-scale flows are
omitted, rapidly rotating convection simulations, and models that
endeavour to reproduce the Sun the best possible way permitted by the
available resources. We consider each of these cases separately.

\subsection{Idealised forced turbulence simulations}
It is often useful to study highly idealised systems where the
turbulence is driven by a body force instead of stratified convection,
and where different physics can be added or removed by hand. In such
setups it is possible to test simple ideas and to see whether the
predictions from mean-field theory can be realised, although such
simulations may have a limited applicability to the real Sun. Another
advantageous aspect of such models is that by virtue of their
simplicity, they are much easier to control and analyze than
simulations with convection.

One example of such idealised simulations, reproducing results of
mean-field theory and supplying possible hints as to how the Sun is
working was recently reported by Mitra et al.\ (2010). In this model,
helically forced turbulence produces an $\alpha$-profile that changes
sign at the equator. Since large-scale flows are omitted, this system
can only host an $\alpha^2$-dynamo. Given that the forcing is
sufficiently helical, a dynamo which produces large-scale magnetic
fields, is excited. The remarkable aspect of this dynamo is that it
produces equatorward migrating active regions in the absence of
shear. Such configurations have previously been obtained only in
mean-field models of $\alpha^2$-dynamos (e.g.\ Baryshnikova \&
Shukurov 1987; R\"adler \& Br\"auer 1987). Although this process is an
unlikely main driver of equatorward migration in the Sun, it may still
contribute to the observed activity.

\subsection{Simulations of rapidly rotating stars}
Photometric observations suggest that stars rotating even much faster
than the Sun are likely to have a comparable absolute differential
rotation $\Delta\Omega=\Omega_{\rm equator}-\Omega_{\rm pole}$ (e.g.\
Korpi \& Tuominen 2003 and references therein). On the other hand, the
magnitude of the $\alpha$-effect is, to first order, proportional to
the rotation rate. Furthermore, test-field simulations indicate that
turbulent diffusion decreases as a function of rotation (K\"apyl\"a et
al.\ 2009b). Combined, these very crude estimates suggest that the
dynamo number, proportional to $\alpha \Delta \Omega$, is greater 
in a rapidly rotating star than in the Sun. At face value this seems to
indicate that exciting a turbulent large-scale dynamo in a simulation
with faster than solar rotation should be easier than with solar
values.

There are some indications that lend credence to this conjecture, see
e.g.\ the studies of Brown et al.\ (2007, 2010) and K\"apyl\"a et al.\
(2010c). These simulations exhibit a solar-like rotation profile with
a fast equator and slow poles but also non-axissymmetric nests of
convection near the equator (e.g.\ Busse 2002; Brown et al.\ 2008)
which are not observed at least in the Sun. Meridional flows are
concentrated in a number of small cells. The radial gradient of
$\Omega$ near the equator is positive, whereas at higher latitudes
differential rotation is much weaker. Negative (positive) kinetic
helicity in the Northern (Southern) hemisphere suggests a positive
(negative) $\alpha$-effect (e.g.\ K\"apyl\"a et al.\ 2010c). According
to mean-field theory a poleward propagating dynamo wave should appear
in regions close to the equator, which is indeed realized in the
simulations. However, the $\alpha$-effect has not been measured
directly in any of the studies reported so far. The large-scale
magnetic fields tend to fill most of the convection zone suggesting
that a distributed dynamo is at operation. If an overshoot layer is
included, a non-oscillating field resides in the stable layer below
the convection zone (K\"apyl\"a et al.\ 2010c). The large-scale
magnetic field is also to a fairly high degree axissymmetric (e.g.\
Brown et al.\ 2010).

Although increasing the rotation rate above the solar value makes it
easier to excite a dynamo in the simulations, other issues arise:
observational studies suggest that the magnetic activity of rapidly
rotating stars appears in the form of strong non-axissymmetric
structures at very high latitudes (e.g.\ Berdyugina \& Tuominen 1998),
which possibly show similar magnetic cycles as the Sun (e.g.\ Jetsu et
al.\ 1993) in which the high-latitude spots alternate in
strength. Such configurations can be obtained from mean-field models
(e.g.\ Elstner \& Korhonen 2005) but not in direct simulations as of
yet.

\subsection{Solar simulations}
The first successful convection-driven dynamo simulations producing
large-scale magnetic fields were performed already in the eighties 
by Gilman (1983)
and Glatzmaier (1985). The rotation profile of these simulations was
solar-like, i.e.\ equator rotating faster than the poles, with a
positive radial gradient of $\Omega$ was found near the equator. Thus
the activity belts migrated towards the poles in contradiction to the
Sun. Although these studies demonstrated the possbility of large-scale
dynamo action, many parameters were not exatcly solar-like. For
example, Gilman (1983) used the Boussinesq approximation and a
rotation rate that is likely greater than in the Sun.

Later studies have refined these models further, using the anelastic
approximation in conjunction with thermal stratification computed from
solar structure models, and accurate physical parameters such as the
solar rotation rate and luminosity (e.g.\ Elliott et al.\ 1999; Brun
\& Toomre 2002; Brun et al.\ 2004; Miesch et al.\ 2000,2006,2008) but
often omitting an overshoot layer and the shear layers near the
surface and at the bottom of the solar convection zone.

Majority of the studies quoted above present hydrodynamical
simulations which concentrate on the study of differential
rotation. The problem there is that even for solar rotation rates the
resulting rotation profile is dominated by the Taylor--Proudman
balance leading to cylindrical isocontours of $\Omega$ (e.g.\ Brun \&
Toomre 2002). This is the so-called `Taylor number puzzle' encountered
earlier in mean-field models (e.g.\ Brandenburg et al.\ 1991). It
turns out that reproducing the solar interior rotation
self-consistently is very hard indeed and none of the current
simulations are able to do this. The solutions are also sensitive
to boundary conditions: fixing the energy flux on the outer boundary
appears to alleviate the Taylor--Proudman constraint (Elliott et al.\
1999). Furthermore, imposing a temperature difference of around $10$~K
on the lower boundary leads to a thermal wind contribution that has a
similar effect (Miesch et al.\ 2006). Such temperature gradients occur
in the simulations naturally but they are apparently not strong
enough. However, as the imposed latitudinal variation of temperature
is transmitted to the convection zone by thermal diffusivity, the
efficiency of the forcing diminishes as the resolution is increased
and diffusivity lowered (e.g.\ Miesch et al.\ 2008). Thus a more robust
method for sustaining a latitudinal temperature gradient is likely to
be needed.

The dynamo action of such solar simulations was studied by Brun et
al.\ (2004). However, in their simulations no appreciable large-scale
magnetic fields were found: a strong small-scale magnetic field is 
obtained but
the large-scale field is only of the order of a few per cent of the
total. This suggests that a fluctuation dynamo is excited but the
large-scale dynamo is subcritical or suppressed. These results are
puzzling, given that the simulations claim to use real solar
parameters. Taken at face value the results suggest that there is no
turbulent mean-field dynamo within the solar convection zone. However,
a number of ingredients are still missing, the most important of which
concern the rotation profiles realised in the simulations.

More specifically, it is currently not possible to reproduce the
surface shear layer or the tachocline at the base of the solar
convection zone with direct numerical simulations. Especially the
tachocline has been considered to host the solar dynamo. Introducing a
tachocline by hand does indeed enable a large-scale dynamo (Browning
et al.\ 2006), although the field is mostly confined in the overshoot
layer and does not show reversals of polarity. In a more recent study,
a similar model did show oscillatory behaviour (Ghizaru et al.\
2010). It is not clear why the latter shows oscillations while the
former does not, although it is possible that the earlier simulation
was simply not ran long enough. However, even in the study of Ghizaru
et al.\ (2010) the activity is at too high latitudes and the migration
of the activity belts towards the equator is not very pronounced.

\section{Possible missing ingredients}
\label{sec:missing}

\subsection{Insufficient resolution}
The most obvious defect of all current simulations is the lack of
numerical resolution. This is also the most difficult problem to solve
because ultimately only bigger computers and codes that scale well in
them give a proper solution. However, in the meantime the current
setups need to be optimised to take full advantage of the resources
available. This includes increasing the resolution near the surface
where the pressure scale height is small by means of a non-uniform
grid.

A more radical solution is to omit certain parts of the star in the
models in order to increase the resolution in the remaining
areas. This is motivated by the fact that the large-scale magnetic
activity in the Sun is concentrated near the equator and that the
sunspots appear more or less independent of longitude. These
observations suggest that the essential ingredients of the solar
dynamo could be captured by modelling only the relevant latitudes and
a reduced longitudinal extent. Simulations in such `wedge' geometry
have been used in the past (e.g.\ Robinson \& Chan 2001; DeRosa \&
Hurlburt 2003) and more recent convection simulations (K\"apyl\"a et
al.\ 2010c,d) seem to compare well with results in full spheres (e.g.\
Brown et al.\ 2010).

\subsection{Unresolved effects of turbulence}
It is clear from Table~\ref{tab:dimpa} that it is not possible to
resolve all the physically relavant turbulent scales in current
simulations. It is also not obvious what effect these scales would
have on the resolved scales. However, it is in principle possible to
take these effects into account by applying suitable subgrid-scale
models. Furthermore, the subgrid-scale models need to be validated by
comparing their results with local numerical simulations (e.g.\
Snellman et al.\ 2009; Garaud et al.\ 2010).

For example, the Taylor--Proudman balance could be alleviated if the
anisotropy of turbulent heat transport due to rotation is taken into
account (e.g.\ Kitchatinov et al.\ 1994). This effect has been
successfully used in hydrodynamical mean-field models (e.g.\ Durney \&
Roxburgh 1971; Brandenburg et al.\ 1992; R\"udiger et al.\ 2005;
K\"uker \& R\"udiger 2008) but not so far in three dimensional
simulations. Similar modelling could be adopted for the Reynolds
stress and electromotive force as well. In particular the
non-diffusive part of the Reynolds stress, often referred to as the
$\Lambda$-effect (e.g.\ R\"udiger 1989), is likely to be important in
sustaining the surface shear layer. On the other hand, the lack of
large-scale magnetic field in solar simulation without overshoot (Brun
et al.\ 2004) could indicate that the relevant scale for the
$\alpha$-effect is not resolved which could be remedied by introducing
it via a subgrid-scale model.

\subsection{Surface shear layer}
The idea that the solar dynamo resides close to the surface arises
from sunspot observations which are consistent with the picture that
the spots are initially formed at a depth of around $r=0.95R_\odot$,
which coincides with the lower part of the surface shear layer. The
surface dynamo idea has recently been revived in the paper of
Brandenburg (2005) who demonstrated that non-helical turbulence with
radial and latitudinal shear leads to a large-scale dynamo and
structures that resemble bipolar regions. In the Sun an
$\alpha$-effect is also likely to be present in these depths so
conceivably an oscillatory dynamo could be obtained or the direction
of the dynamo wave reversed near the surface (e.g.\ K\"apyl\"a et al.\
2006b).

None of the current solar simulations, however, capture the surface
shear layer even in the most stratified and highest resolution runs
performed so far (Miesch et al.\ 2008). This can be due to still
insufficient density stratification and scale sepration rendering the
$\Lambda$-effect ineffective near the surface. Another possible reason
is that shear layer is sensitive to the outer boundary condition of
the simulations.

\subsection{Tachocline}
The studies of Browning et al.\ (2006) and Ghizaru et al.\ (2010) have
highlighted the importance of the tachocline for the dynamo. The
problem here is that it is not yet possible to form a tachocline
self-consistently in the simulations so it has to be enforced by some
method. This is because the diffusivities in the current simulations
are still so large that the differential rotation from the convection
zone diffuses into the stable layer. This could be countered by
lowering the diffusion coefficients below the convection zone but this
is likely to cause numerical issues. Another issue is the stability of
the tachocline with respect to magnetic fields: while certain dynamo
models assume that the field in the tachocline needs to be of the
order of $10^5$~Gauss (e.g.\ Dikpati \& Charbonneau 1999), other
studies indicate that the tachocline shear becomes unstable already for
field strenghts that are two orders of magnitude lower (e.g.\ Arlt et
al.\ 2005).

\subsection{Meridional circulation}
Another large-scale flow component in the Sun is the meridional flow.
Observations indicate that the flow is poleward at the surface with a
magnitude of around 10~m~s$^{-1}$. Hydrodynamic mean-field models
indicate that the flow consists of a single counter-clockwise cell
(e.g.\ Rempel 2005). The return flow should reside near the base of
the convection zone with a magnitude of 1~m~s$^{-1}$. If the solar
dynamo also resided in the deep layers of the convection zone, the
deep return flow could in principle change the direction of the dynamo
wave. This flow configuration would also be essential for the
flux-transport dynamo model to work (e.g.\ Dikpati \& Charbonneau
1999). However, current simulations do not show a clear single cell
configuration but rather a large number of smaller cells (e.g.\ Brun
et al.\ 2004). Furthermore, helioseismology is currently unable to
provide observational confirmation of the structure of the meridional
flow within the solar convection zone.

\section{Conclusions}
\label{sec:discuss}
Current global simulations of solar magnetism struggle with several
aspects of the observed activity: firstly, it is difficult to obtain a
large-scale field at all without a tachocline (Brun et al.\
2004). Secondly, when a tachocline is imposed, there is a large-scale
field but it is non-oscillating (Browning et al.\ 2006) or the
activity belts are at too high latitudes (Ghizaru et al.\ 2010). Most
of the problems can be associated with insufficient numerical
resolution due to which some of the physically relevant scales of
turbulence are not resolved, and prevents the self-consistent
generation of a tachocline and the surface shear layer.

Relatively little can be done to overcome the resolution issue,
although simulations in the `wedge' geometry (Mitra et al.\ 2009,
2010; K\"apyl\"a et al.\ 2010c,d) promise to deliver some
benefits. However, a perhaps more promising alternative is to
introduce improved subgrid-scale models, capturing also non-diffusive
effects of turbulence, into the simulations. However, this is also
likely to be a long and rocky road because the subgrid-scale models
should be validated as rigorously as possible before their use.

Furthermore, our current understanding of the existing simulations
that are capable of large-scale dynamo action (e.g.\ Browning et al.\
2006; Brown et al.\ 2010; K\"apyl\"a et al.\ 2010c; Ghizaru et al.\
2010) is still quite insufficient. For example, there are only
enlightened guesses of the $\alpha$-effect based on the sign of
kinetic helicity, and even less is known of turbulent transport
coefficients related to other dynamo mechanisms such as the
$\meanv\Omega\times\meanv{J}$ and shear-current effects or turbulent
diffusivity. Therefore the current simulations should be analyzed in
greater detail, e.g.\ with the help of test-field methods and
corresponding mean-field models, in order to find out which effects
are responsible for the dynamo and where it is situated. Such analysis
could also help to understand what is missing from the simulations in
comparison to the Sun.

\acknowledgements{Computational resources granted by CSC -- IT Center
  for Science, who are financed by the Ministry of Education, and
  financial support from the Academy of Finland grants No.\ 121431 and
  136189 are acknowledged.}


\end{document}